# Three-dimensional acoustic double-zero-index medium with a Dirac-like point

Changqing Xu[1], Guancong Ma[2*], Ze-Guo Chen[1], Jie Luo[3,4], Jinjie Shi[4], Yun Lai[3,4*] and Ying Wu[1*]

[1] Division of Computer, Electrical and Mathematical Science and Engineering, King Abdullah University of Science and Technology (KAUST), Thuwal 23955-6900, Saudi Arabia

[2] Department of Physics, Hong Kong Baptist University, Kowloon Tong, Hong Kong

[3] National Laboratory of Solid State Microstructures, School of Physics, and Collaborative Innovation Center of Advanced Microstructures, Nanjing University, Nanjing 210093, China

[4] College of Physics, Optoelectronics and Energy & Collaborative Innovation Center of Suzhou Nano Science and Technology, Soochow University, Suzhou 215006, China

Emails: phgcma@hkbu.edu.hk, laiyun@nju.edu.cn, Ying.Wu@kaust.edu.sa

Abstract:

We report a design and experimental realization of a three-dimensional (3D) acoustic double-zero-index medium (DZIM), whose effective mass density and compressibility are nearly zero simultaneously. The DZIM is constructed from a cubic lattice of three orthogonally-aligned metal rods in air. The combination of lattice symmetry and accidental degeneracy yields a four-fold degenerate point with conical dispersion at the Brillouin zone center, where the material becomes a 3D DZIM. Though occupying a finite volume, the 3D DZIM maintains the wave properties of a "void space" and enables rich applications. For demonstration, we fabricate an acoustic "periscope" by placing the designed 3D DZIM inside a 3D bending waveguide, and observe the unusual wave tunneling effect through this waveguide with undisturbed planar wavefront. Our findings establish a practical route to realize 3D DZIM as an effective acoustic "void space," which offers unprecedented opportunities for advanced sound manipulation.

A wave propagating in a medium with one or more constitutive parameters vanishing does not accumulate any phase retardation. This characteristic can be leveraged for a number of unique wave functionalities such as wavefront engineering[1-10], cloaking objects[8-14], wave tunneling[15-23], asymmetric transmission[24-26] and photonic/phononic doping[27-32]. A single-zero-index medium, with only one constitutive parameter near-zero, usually has a significant impedance mismatch with the background medium, which is undesirable for real applications. A double-zero-index medium (DZIM), with both constitutive parameters near-zero, can overcome this obstacle, owing to its finite-valued effective impedance[7-11]. In the past decade, various approaches have been proposed to realize DZIM for electromagnetic[1-4,6,8,9,11,12,16,17,19-23,27-32] and acoustic waves[5,7,10,13-15,18,24-26]. One route relies on the realization of the Dirac-like conical dispersion at the Brillouin zone center using the accidental degeneracy of states[7-10,18,22]. Another approach is the doping of single-zero-index media[27-32], which is an extraordinary consequence of impurities. A third method of realizing impedance-matching in zero-index medium involves the introduction of parity-time symmetry[33]. However, to the best of our knowledge, DZIM has only been realized in two-dimensional (2D) systems so far, which restrict their applications to in-plane propagating waves in planar structures[5,13]. Due to the complexity induced by higher dimensions, both the structural design and experimental realization in 3D systems are more difficult than those in 2D systems. For airborne acoustics, they are fundamentally more challenging because of the big contrast in mass density and speed of sound between air and solids.

In this work, we demonstrate the design and experimental realization of a 3D acoustic DZIM with effective mass density and compressibility approaching zero simultaneously. The design principle is based on the combination of lattice symmetry and accidental degeneracy. The 3D DZIM is a phononic crystal (PC) consisting of a cubic lattice of unit structures with three orthogonally-aligned metal rods in air, which can be regarded as sound-hard scatterers. With non-symmorphic glide symmetries[34-37] in the three spatial directions, the lattice symmetry enables a three-fold degeneracy at the Brillouin Zone center corresponding to three dipolar resonances. Moreover, accidental degeneracy can be further achieved by tuning the geometry of the unit structure such that the monopolar resonance overlaps with the dipolar ones, leading to a four-fold degenerate point. In the vicinity of this point, the band structure manifests linear dispersion in all directions of the reciprocal space, which looks similar to the Weyl point despite a different physical origin. We call it a Dirac-like point in 3D k-space. By using an effective parameter retrieval method[38,39], we affirm that the effective mass density and compressibility simultaneously cross zero at the frequency of the Dirac-like point. By experimentally constructing the 3D acoustic DZIM and putting it inside a bending waveguide with two 90° bends, we demonstrate the unusual wave tunneling functionality with a non-disturbed planar wavefront at the exit. This is a direct proof of the double-zero-index property in three dimensions. To the best of our knowledge, our work is the first realistic theoretical proposal and experimental realization of a 3D DZIM. Although a 3D DZIM occupies a finite volume, it inherits some wave properties of an effective "void space". This rare physical nature leads to many unusual applications such as 3D wave tunneling and cloaking of 3D objects, etc.

**Results**

**Realization of 3D DZIM by using a PC with glide symmetries.** We begin with a PC of a cubic lattice, whose unit cell consists of three orthogonally-aligned aluminum rods in air, as illustrated in Fig. 1a. The lattice constant is $a$. The aluminum rods are aligned in the $x$-, $y$-, $z$-axes, respectively, all having a square cross section with a side length $L$. When the axes of the rods are separated by $a/2$, the system has glide symmetries[34-37], which can be defined as the combination of reflection symmetry and translation by half of a lattice constant:

$$G_{xy} = \left\{ M_{-xy} \mid \frac{a}{2}\hat{z} \right\} : (x, y, z) \to \left( -y, -x, z + \frac{a}{2} \right),$$

$$G_{yz} = \left\{ M_{-yz} \mid \frac{a}{2}\hat{x} \right\} : (x, y, z) \to \left( x + \frac{a}{2}, -z, -y \right),$$

$$G_{zx} = \left\{ M_{-zx} \mid \frac{a}{2}\hat{y} \right\} : (x, y, z) \to \left( -z, y + \frac{a}{2}, -x \right). \quad (1)$$

These non-symmorphic symmetry operators $G_{xy}$, $G_{yz}$, and $G_{zx}$ transform the $i$th ($i=x,y,z$) rods into the $j$th ($j=y,z,x$) rods. The lattice also has $C_3$ rotational symmetry about the [111] diagonal axes[40] and mirror symmetries about the $xy$, $yz$, $zx$ planes, as shown in Figs. 1a-1b.

The band structure of the PC with $L=0.3a$ is shown in Fig. 1c, which exhibits a three-fold degeneracy at the $\Gamma$ point for the second, third and fourth branches. The eigenfields of the three-fold degenerate states and the single state above it (on the fifth branch) are plotted in Figs. 1d-1g, respectively. Clearly, the single state is a monopolar state while the three-fold degenerate states are three dipolar states. The degeneracy of the dipolar states at the $\Gamma$ point is guaranteed by the symmetry of the system, regardless of the geometry. Altering the size of the rods only changes the eigenfrequency of these states. Therefore, accidental degeneracy of the monopolar and dipolar states can be further achieved by altering the parameter L.

The band structure of a PC with $L=0.35a$ is plotted in Fig. 2a. In this case, the monopolar state overlaps with the three-fold degenerate dipolar states, leading to a four-fold degenerate point at the Γ point. An enlarged view of the band structure near the Γ point in the $k_x - k_y$ plane is shown in Fig. 2b, in which a conical dispersion intersecting with two flat bands at the four-fold degenerate point is observed. By examining the dispersions in the $k_y - k_z$ and $k_z - k_x$ planes, we observe almost the same conical dispersion as well as the intersecting flat bands (see Supplementary Figure 2 for details). Therefore, Dirac-like linear dispersion emerges in the vicinity of this four-fold degenerate point in all directions. This rare property is similar to that of the Weyl point. However, the physical origins are intrinsically different. A Weyl point is induced by breaking time-reversal symmetry or/and parity inversion[41-43], while the Dirac-like linear dispersion presented here is induced by the accidental degeneracy of eigenstates with different parities[7-10,18,22]. We call this four-fold degenerate point as a Dirac-like point in the 3D k-space.

From a tight-binding model, the Hamiltonian near the Dirac-like point can be described in the basis of s state and p states:

$$H_0 = \begin{pmatrix} E_s(\vec{k}) & -2it_{sp}\sin(k_x) & -2it_{sp}\sin(k_y) & -2it_{sp}\sin(k_z) \\ 2it_{sp}\sin(k_x) & E_{px}(\vec{k}) & 0 & 0 \\ 2it_{sp}\sin(k_y) & 0 & E_{py}(\vec{k}) & 0 \\ 2it_{sp}\sin(k_z) & 0 & 0 & E_{pz}(\vec{k}) \end{pmatrix}.$$

Here

$$E_s(\vec{k}) = \varepsilon_s + 2t_s\left(\cos(k_x) + \cos(k_y) + \cos(k_z)\right),$$

$$E_{px}(\vec{k}) = \varepsilon_{px} + 2t_\sigma \cos(k_x) + 2t_\pi \cos(k_y) + 2t_\pi \cos(k_z),$$

$$E_{py}(\vec{k}) = \varepsilon_{py} + 2t_\pi \cos(k_x) + 2t_\sigma \cos(k_y) + 2t_\pi \cos(k_z),$$

$$E_{pz}(\vec{k}) = \varepsilon_{pz} + 2t_\pi \cos(k_x) + 2t_\pi \cos(k_y) + 2t_\sigma \cos(k_z), \quad (2)$$

where $\varepsilon_s$, $\varepsilon_{pi}$ (i=x,y,z) are the on-site energy, $k_x\hat{x}$, $k_y\hat{y}$ and $k_z\hat{z}$ are the components of wave vector $\vec{k}$, $t_\sigma$ and $t_\pi$ represent nearest neighbor hoppings. Dirac-like linear dispersion and two flat branches can be immediately obtained in all directions as a consequence of the accidental degeneracy of $E_s=E_{pi}$. (see Supplementary Note 2). The linear dispersion at the Dirac-like point can be exploited for a myriad of fascinating phenomena such as negative refraction[9], Klein tunneling[9,44] and super-collimation[9,44].

On the other hand, the Dirac-like linear dispersion may also appear as a consequence of double zero parameters in DZIM. Suppose the effective mass density and compressibility are both dispersive in frequency and cross zero at frequencies $\omega_d$ and $\omega_m$, respectively. A frequency dependent function can be expanded as follows:

$$f(\omega) = f_{\omega_0} + \frac{\partial f}{\partial \omega}(\omega-\omega_0) + o(\omega^2), \tag{3}$$

where $o(\omega^2)$ represents higher orders. The effective mass density $\rho_{eff}$ can be written as $\rho_{eff} \sim c_\rho(\omega-\omega_d)$ near $\omega_d$, while the effective compressibility $\beta_{eff}$ can be written as $\beta_{eff} \sim c_\beta(\omega-\omega_m)$ near $\omega_m$. Here $c_\rho, c_\beta$ denote the linear coefficients of $\rho_{eff}$ and $\beta_{eff}$, respectively. If the effective mass density and compressibility cross zero simultaneously, i.e., $\omega_d = \omega_m = \omega_D$, by substituting the above formulas into the general formula of dispersion relation, $\omega^2 = c^2 k^2$, in which $c^2 = \frac{1}{\rho\beta} = \frac{1}{c_\rho c_\beta (\omega-\omega_D)^2}$, we obtain the dispersion relation near $\omega_0$ as $k^2 \sim c_\alpha c_\beta \omega_D^2 (\omega-\omega_D)^2$ near $\omega_D$, which denotes Dirac-like linear dispersions near the $\Gamma$ point.

To calculate the effective mass density $\rho_{eff}$ and compressibility $\beta_{eff}$ for our proposed system, we adopt the effective parameter retrieval method based on field averaging[38,39]. The results are plotted in Fig. 2c, which shows the effective mass density $\rho_{eff}/\rho_0$ (solid black line) and compressibility $\beta_{eff}/\beta_0$ (solid blue line) indeed cross zero simultaneously at the frequency of the Dirac-like point $\omega_D = 0.807(2\pi c/a)$ for the case of $L=0.35a$. This confirms that such a PC behaves effectively as the desired DZIM. In comparison, when $L=0.3a$, the effective mass density (black dashed line) and compressibility (blue dashed line) cross zero at different frequencies $\omega_d = 0.82(2\pi c/a)$ and $\omega_m = 0.88(2\pi c/a)$, which correspond to the eigenfrequencies of the dipolar and monopolar states at the $\Gamma$ point in the band structure, respectively. Therefore, only single-zero-index media can be realized in this case. When the band gap between $\omega_d$ and $\omega_m$ closes, the Dirac-like conical dispersion emerges as a result of the double-zero-index. The Hamiltonian and effective medium pictures are consistent with each other.

Besides the Dirac-like linear dispersion, the Dirac-like point is accompanied by the existence of two additional flat bands, as shown in Figs. 2a and 2b. These two flat bands correspond to transverse acoustic waves in which the particle displacement directions are perpendicular to the propagation direction. This unusual characteristic is manifested in the eigenstates plotted in Figs 1f and 1g. Counterintuitive as it may sound, such transverse acoustic waves indeed arise as a direct result of the zero effective mass density (See Supplementary Note 4). Such transverse acoustic waves cannot be easily coupled to normal acoustic waves as longitudinal waves, and therefore their existence does not affect the wave properties near the Dirac-like point. The wave behavior near the Dirac-like point is mainly determined by the linear branches.

With $\rho_{eff}$ and $\beta_{eff}$ both vanishing at the Dirac-like point, the 3D DZIM can be regarded as an acoustic "void space". To demonstrate the unique consequence of this property, we begin by considering the transmission of a normally incident wave through a slab of 3D DZIM with a finite thickness. It is easy to see that the transmission coefficient is always unity, despite the potential impedance mismatch between the DZIM and the background. The DZIM seemingly connects the input surface to the output, thus the space it occupies becomes a void for the wave. We verify this property analytically using the effective medium, and numerically using the PC structure (See Supplementary Note 5, 6). We emphasize that this property is unique to 3D DZIM and cannot be found in any other acoustic materials. In comparison, a wave is always reflected by a slab of an ordinary medium with a different impedance from the background, unless at the frequencies of Fabry-Perot resonances. For a wave incident on a slab of single-zero-index medium, the transmission coefficient decreases when the slab thickness increases (See Supplementary Note 6 and Supplementary Fig. 5 for a demonstration). On the other hand, a DZIM in 2D is only equivalent to "void plane," and its functionalities are limited to in-plane propagating waves[5-10,13]. Next, extending from the transmission problem of a slab, it is straightforward to see that waves can perfectly propagate through a 3D DZIM with a finite volume. For example, consider an incoming wave that is normally incident onto one surface of a cube of 3D DZIM, the incident wave is partitioned into five waves, each carrying an identical amount of energy, that radiate normally from the other 5 surfaces. If we close any 4 outlet surfaces by blocking them with totally reflective boundaries, the DZIM cube becomes a wave-steering device that can direct waves to orthogonal directions in all three dimensions with 100% efficiency.

**Wave tunneling enabled by DZIM.** To exploit the unique property of the 3D DZIM as a 3D

acoustic "void space," we have designed and fabricated a 3D "periscope" for sound. As illustrated in Fig. 3a, it is an acoustic bending waveguide with two 90° bends. The incident wave comes from the top along the negative $z$-direction. The first bend turns the waveguide from negative $z$-direction to negative $y$-direction, and the second bend turns from negative $y$-direction to positive $x$-direction. The waveguide is filled with the PC as 3D DZIM. Figures 3b and 3c show the simulated pressure field distribution with the filling material being the PC and its effective medium, respectively. For both cases, it is seen that the systems have minimal reflection. Moreover, the outgoing waves retain a planar wavefront, but the wavefronts are re-oriented and parallel to the $yz$ plane now. When the DZIM is absent, the wave is inevitably scattered by the bends which severely scrambles the outgoing wavefront, as shown in Fig. 3d. These results clearly demonstrate the equivalence of our PC and a 3D DZIM by wave-tunneling through 3D bending waveguide with wavefront-preserving re-orientation, which is the direct and unique consequence of the 3D DZIM as an effective acoustic "void space."

**Experimental validation.** We build the "periscope" waveguide with 12 aluminum plates. The waveguide is shown in Fig. 4a. It has a square cross-section with a side length of 14.25 cm (5 unit cells). As shown in Fig. 4b, the PC is built inside the waveguide using aluminum rods with specific lengths. The rods have a square cross-section with a side length of 1.0 cm. The lattice constant is 2.85 cm. In the simulation, this PC has a Dirac-like point at $f_{Dirac} = \omega_D / (2\pi) = 9,742 Hz$. Note that to ensure the dipolar states are free from deformation, the unit cell is chosen in the way shown in Fig. 1b, so that the sound-hard PC-air interface will not break the mirror symmetry of the PC. We mount a loudspeaker with a diameter of *12cm* on one end of the waveguide to generate a plane wave that propagates in the negative $z$-direction, as shown in Fig. 4a. The waveguide then bends to negative $y$--direction, then positive $x$-direction. We use a microphone to measure the transmitted wave at the exit of the waveguide. The microphone is mounted on a translational stage to raster-scan the wave profile in the $yz$-plane, the $xy$-plane, and the $xz$-plane.

We perform a 2D Fourier transform on the spatial map in frequencies near $f_{Dirac}$. The Fourier transforms of the $yz$-plane, which is parallel to the PC-air interface at the exit of the waveguide, are shown in Fig. 5a. Near $f_{Dirac}$=9 680$Hz$, the magnitude of the Fourier transform has a sharp peak at $k_y = k_z = 0$. This implies that the acoustic wave exiting from the PC has a wavevector dominated by $k_x$, which indicates a planar wavefront with minimal distortion, despite having changed its propagation direction twice. For frequencies above and below $f_{Dirac}$, the outgoing waves possess $k_y$ and $k_z$ components, as the sharp peak gradually expands into a circle (Fig. 5a). We further examine the 2D Fourier transforms of the rasterized map in the *xy*- and *xz*-planes, as shown in Fig. 5b, c. At 9,680 Hz, the magnitudes sharply peak at $k_x = 2\pi f_{Dirac}/c \approx 177 m^{-1}$, which clearly means that the outgoing wave is a plane wave. When the frequency deviates from $f_{Dirac}$, the wave starts to possess $k_y$ and $k_z$ components. Also, the results obtained at 9,800 Hz and 9,560 Hz are noticeably noisier than at $f_{Dirac}$, which indicates a reduction in the transmission coefficient when the frequency deviates from $f_{Dirac}$. This is expected since the PC is a DZIM at $f_{Dirac}$. In Fig. 5d and Fig. 5e, we plot the real-space map of the pressure fields at $f_{Dirac}$ in the *xy*- and *xz*-planes. Despite the presence of some noise, which is unavoidable due to the non-ideal acoustic condition of our laboratory, well-defined planar wavefronts can be clearly identified, which unambiguously shows that the sound remains a plane wave after bending twice in the waveguide. These experimental results convincingly show that at normal incidence, a plane wave at $f_{Dirac}$ can go through the PC with minimal distortion, and can be re-directed into another direction that is parallel to the PC-air interface. This effect is the signature of a DZIM. We note that the measured $f_{Dirac}$ is 9,680 Hz, which deviates from the prediction by only ~0.4% and is well within the fabrication tolerance.

The 3D sound tunneling effect is not affected by the presence of sound-hard defects. To show this, we place a defect with the shape and size of one unit cell inside the waveguide. The defect has sound-hard boundaries. The results are shown in the Supplementary Note 7. It can be seen that when the defect is embedded in the PC, it generates almost no scattering, and the outgoing wavefront remains intact.

**Conclusions**

In conclusion, we have provided a theoretical recipe for design 3D DZIM using PCs. Due to the combination of lattice symmetries and the accidental degeneracy, the PC possesses a Dirac-like point at the Brillouin zone center in 3D k-space. The linear dispersion is affirmed by its Hamiltonian from a tight-binding model. On the other hand, by using an effective parameter retrieval method, we show that the effective mass density and compressibility of the PC approaching zero simultaneously at the frequency of Dirac-like point. This leads to near zero effective refractive index and finite-valued effective impedance in all three spatial directions. Such rare property bestows the PC some intriguing functionalities of an effective acoustic "void space," including sound tunneling through an arbitrarily-shaped 3D waveguide with high transmittance. Based on this recipe, we have designed and experimental realized a 3D DZIM. We demonstrate the effectiveness of our approach and the property of the DZIM by an acoustic "periscope" bending waveguide filled with the PC. Experimental results conclusively show that the sound can tunnel through the bends while maintaining a planar wavefront, which is an important characteristic of the DZIM. Our findings not only demonstrate that the concept of DZIM can be extended to three dimensions, but also provide a novel platform for advanced 3D sound manipulation.

**Methods**

**Simulations.** The band structures, eigenstates (Figs. 1c-g, 2a, 2b) and field distributions (Figs. 3b-d) are calculated by the acoustic module in commercial Finite-Element method software (COMSOL MULTIPHYSICS). The effective mass density and compressibility (Fig. 2c) are calculated by the field averaging at the surface of the unit cell in COMSOL MULTIPHYSICS. The density and velocity of sound in air are chosen to be 1.21 kg/m$^3$ and 343m/s, respectively. The three orthogonally-aligned aluminum rods are treated as rods with hard-wall boundaries because of the huge impedance mismatch between air and aluminum.

**Acknowledgements**

The work described in here is partially supported by King Abdullah University of Science and Technology (KAUST) Office of Sponsored Research (OSR) under Award No. OSR-2016-CRG5-2950 and KAUST Baseline Research Fund BAS/1/1626-01-01. G. M. is supported by the Hong Kong Research Grants Council (grant no. RGC-ECS 22302718, and ANR-RGC A-HKUST601/18, CRF C6013-18GF), the National Natural Science Foundation of China (grant no. 11802256), and by the Hong Kong Baptist University through FRG2/17-18/056. Y. L. is supported by National Key R&D Program of China (2017YFA0303702), National Natural Science Foundation of China (Grant No. 61671314 and No. 11634005).

**Author contributions**

C. X. designed the phononic crystal and performed the numerical simulations. G. M. designed the experiment, and carried out the measurements and data analysis. G. M. and J. S. assembled the "periscope" waveguide and set up the experiment. Z.-G. C., C. X. and Y. W. contributed to the theoretical analysis of Hamiltonian. J. L. and Y. L. provided the effective parameter retrieval method. C. X., G. M., Y. L. and Y. W. wrote the manuscript with inputs from all authors. The project was supervised by G. M., Y. L. and Y. W.

**Competing interests:** The authors declare no competing interests.

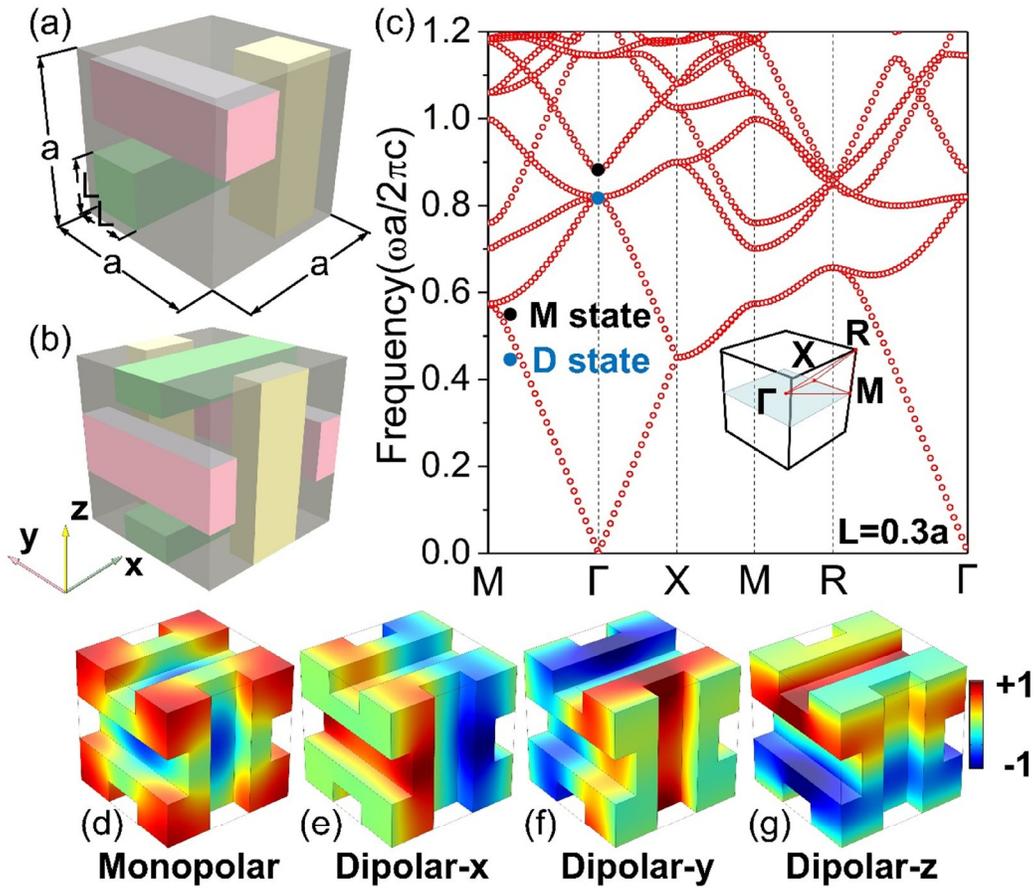

**Fig. 1** Band structure and eigenstates of the 3D PC with glide symmetry. **a, b** Different types of unit cell selection for the same PC. The lattice constant is *a* and the host medium (gray) is air. Colored square blocks with side length $L \times L \times a$ are aluminum blocks. **c** Band structure of the PC with $L=0.3a$. Black and blue dots represent the monopolar state and the three-fold degenerate dipolar states at the Brillouin Zone center. **d-g** Acoustic pressure field distributions of the monopolar state and dipolar states.

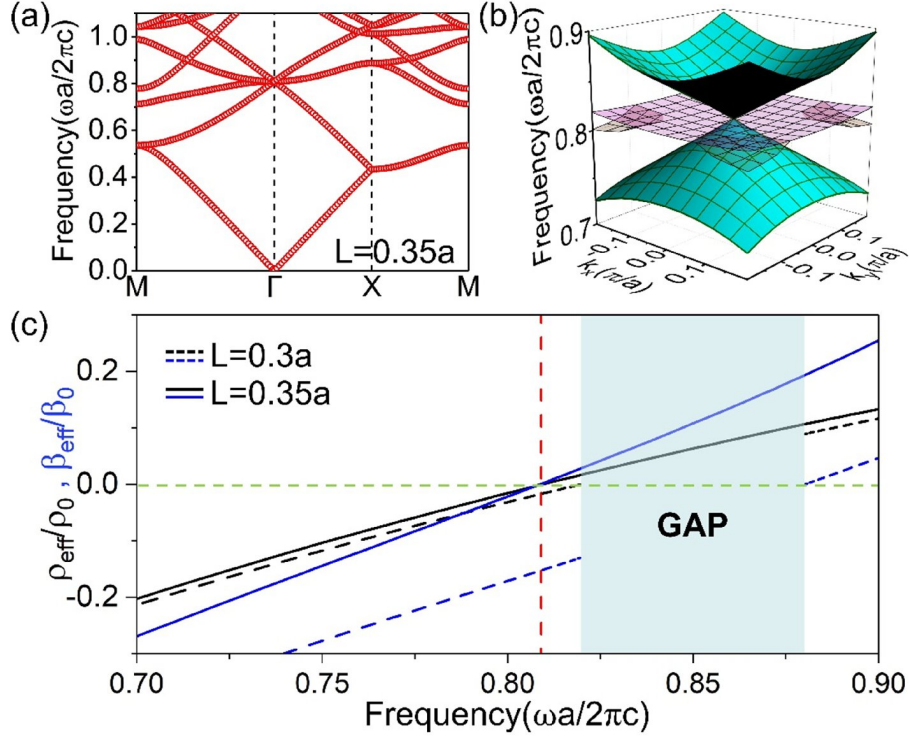

**Fig. 2** Realization of 3D acoustic DZIM by accidental degeneracy. **a** Band structure of the PC with $L=0.35a$, which shows a four-fold degenerate point (the Dirac-like point) with a conical dispersion in its vicinity at the $\Gamma$ point. **b** A zoom-in plot of the conical dispersion surfaces near the Dirac-like point in the $k_x - k_y$ plane. **c** The normalized effective mass density $\rho_{eff}$ (black curves) and compressibility $\beta_{eff}$ (blue curves) for the case of $L=0.35a$ (solid curves) and $L=0.3a$ (dashed curves). For the PC with $L=0.35a$, the effective parameters cross zero simultaneously at the frequency of the Dirac-like point ($\omega_D = 0.807(2\pi c/a)$). While for the PC with $L=0.3a$, the effective parameters cross zero at different frequencies (frequencies of *M* and *D* states shown in Fig. 1c) and only single-zero-index medium is obtained.

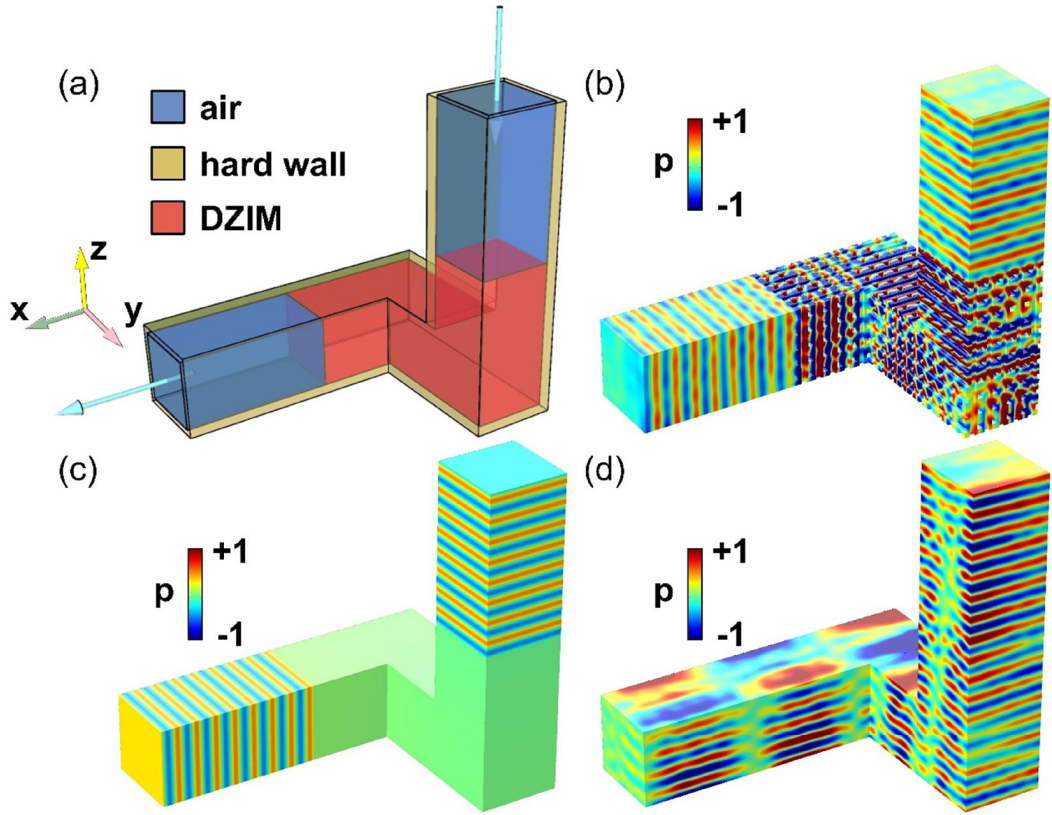

**Fig. 3** Transmission of sound through an acoustic "periscope" filled with DZIM. **a** Illustration of the designed "periscope" for sound, which is a 3D bending waveguide with two 90° bends (gold color). The red region represents the 3D DZIM and the blue region represents air. Waves can perfectly tunnel through the 3D bending waveguide filled with either the designed PC as the 3D DZIM (**b**) or the 3D DZIM with effective parameters of the PC (**c**), respectively. The planar wavefront is well preserved at the exit. **d** In comparison, the wave is inevitably scattered by the bends without the 3D DZIM filling, which severely scrambles the outgoing wavefront.

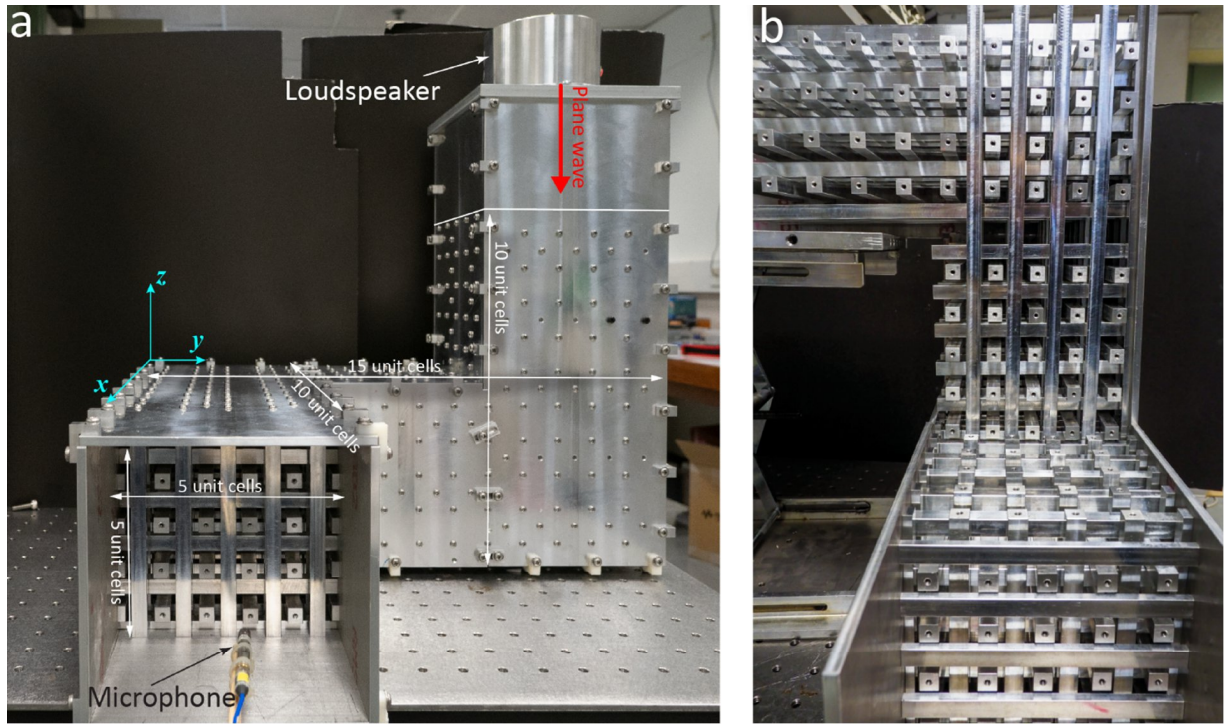

**Fig. 4** Experimental setup. **a** A loudspeaker emits a plane wave in the negative *z*-direction. The waveguide bends twice and the output is in the positive *x*-direction. A microphone is mounted on a translational stage to scan the sound in the output side in all *yz*-, *xy*-, and *xz*-planes. **b** Some boundaries of the waveguide are removed to show the PC inside.

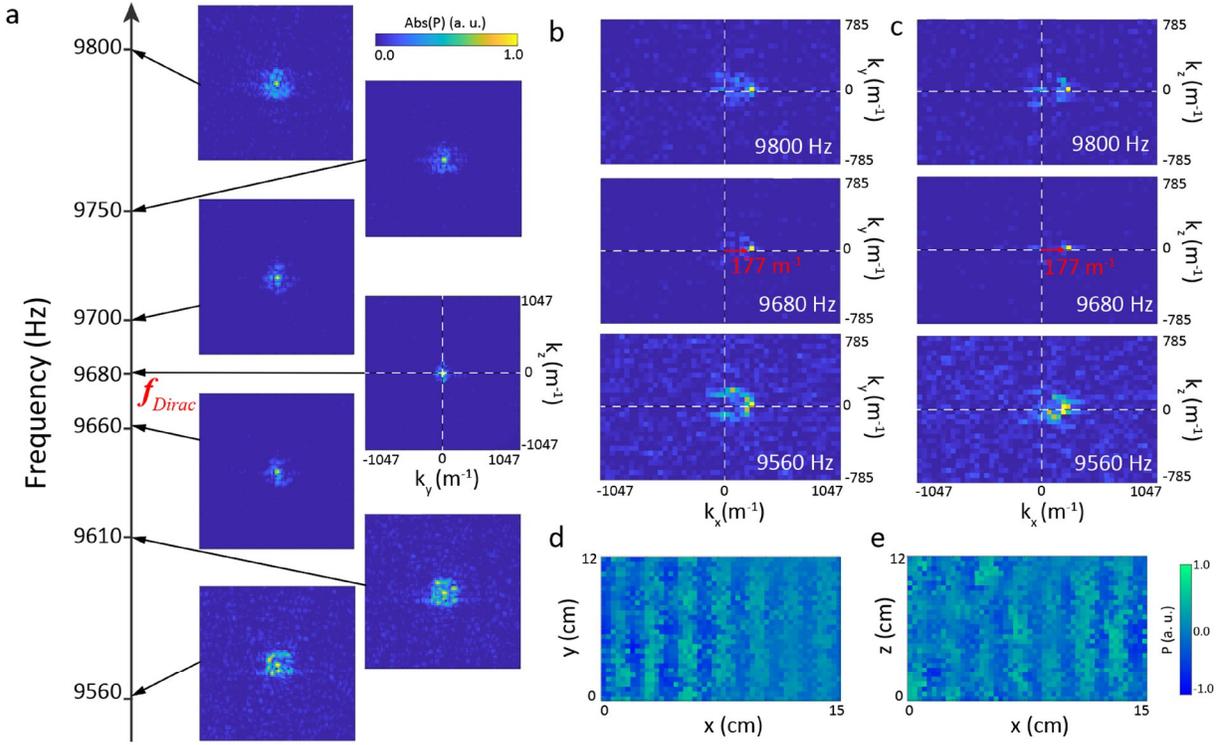

**Fig. 5** Plane wave tunneling with minimal distortion. **a** The 2D Fourier transforms of the scanned field patterns near the frequency of the 3D Dirac-like point $f_{Dirac} = 9,680 Hz$. In the $k_y - k_z$ plane, the Fourier transform has a distribution that peaks at $k_y = k_z = 0$ at $f_{Dirac}$, the distribution gradually expands as the frequency moves away from $f_{Dirac} = 9,680 Hz$. All maps in **a** have the same coordinates. **b,c** In both the $k_x - k_y$ and $k_x - k_z$ planes at $f_{Dirac}$ (middle panels), the Fourier transforms imply a plane wave in the positive *x*-direction. The red arrows mark the position of $k_x = 2\pi f_{Dirac}/c = 177 m^{-1}$, $k_y(k_z) = 0$. Away from $f_{Dirac}$, non-zero $k_y(k_z)$ components start to appear. **d** and **e** show the real-space maps of *xy*-plane and *xz*-plane at $f_{Dirac}$ respectively, wherein planar wavefront can be clearly seen.

18